\documentclass{epl}

\title{Nonextensive diffusion as nonlinear response}

\author{James F. Lutsko \inst{1} \and
Jean Pierre Boon \inst{1} }
\institute{
  \inst{1} Center for Nonlinear Phenomena and Complex Systems\
       Universit\'e Libre de Bruxelles, 1050 - Bruxelles, Belgium
       (E-mail: {\tt jlutsko@ulb.ac.be}; {\tt jpboon@ulb.ac.be})
}

\usepackage{amsmath}

\begin{document}

\maketitle

\begin{abstract}
The porous media equation has been proposed as a phenomenological
``non-extensive'' generalization of classical diffusion. Here, we show that
a very similar equation can be derived, in a systematic manner, for a
classical fluid by assuming nonlinear response, i.e. that the diffusive flux
depends on gradients of a power of the concentration. The present equation
distinguishes from the porous media equation in that it describes \emph{%
generalized classical} diffusion, i.e. with $r/\sqrt Dt$ scaling, but 
with a generalized Einstein relation, and with power-law probability
distributions typical of nonextensive statistical mechanics.
\end{abstract}

One of the characteristic features of the nonextensive thermodynamics
introduced by Tsallis is the appearance of non-exponential distribution
functions with power-law tails \cite{SwinneyPhysicaD, TsallisGellman}. There
has been considerable interest recently in the question of how such
non-exponential distributions might arise from first-principle considerations%
\cite{TsallisPhysicaD, BeckCohen, HanelThurner, AbeThurner}. In this paper,
we show that it is possible to obtain, in a generic manner, power-law
distributions for the diffusion of a tracer particle in a liquid by assuming
a simple generalization of the usual linear response arguments.

In classical transport theory, the probability to find the diffuser at point $%
\overrightarrow{r}$ at time $t$ given that it starts at point $%
\overrightarrow{r}_{0}$ at time $0$ obeys the advection-diffusion equation%
\begin{equation}
\frac{\partial }{\partial t}P\left( \overrightarrow{r},t;\overrightarrow{r}%
_{0},0\right) =\frac{\partial }{\partial \overrightarrow{r}}\cdot 
\overrightarrow{u}\left( \overrightarrow{r},t\right) P\left( \overrightarrow{%
r},t;\overrightarrow{r}_{0},0\right) +\frac{\partial }{\partial 
\overrightarrow{r}}\cdot \overleftrightarrow{D}\cdot \frac{\partial }{%
\partial \overrightarrow{r}}P\left( \overrightarrow{r},t;\overrightarrow{r}%
_{0},0\right) \,,  \label{0}
\end{equation}%
where $\overrightarrow{u}\left( \overrightarrow{r},t\right) $ is the drift
and $\overleftrightarrow{D}$ the diffusion tensor, see e.g. \cite{VanKampen}.
Note that in general, we could, instead of probabilities, speak of the
concentration of a diffusing species or the density of a single-component
system undergoing self-diffusion just as well. There are currently two
widely explored generalizations of this classical diffusion equation. The
first involves a fractional time derivative in the diffusion term 
\cite{MetzlerKlafter}. This fractional diffusion equation can be derived,
e.g., by considering the behavior of a diffuser on a lattice with the
waiting time between hops being governed by a power-law distribution (e.g.,
a Pareto distribution), a type of Levy flight \cite{MetzlerKlafter}. A
second generalization of the classical diffusion equation sometimes used to
model anomalous diffusion is the so-called ``porous media equation"%
\begin{equation}
\frac{\partial }{\partial t}P\left( \overrightarrow{r},t;\overrightarrow{r}%
_{0},0\right) =\frac{\partial }{\partial \overrightarrow{r}}\cdot 
\overrightarrow{u}\left( \overrightarrow{r},t\right) P\left( \overrightarrow{%
r},t;\overrightarrow{r}_{0},0\right) +\frac{\partial }{\partial 
\overrightarrow{r}}\cdot \overleftrightarrow{D}\cdot \frac{\partial }{%
\partial \overrightarrow{r}}P^{\mu }\left( \overrightarrow{r},t;%
\overrightarrow{r}_{0},0\right) \,.  \label{1}
\end{equation}%
As its name indicates, this equation was established (empirically) for 
particular physical systems \cite{Muskat}, and, recently, its formal structure,
$\partial_t f = \Delta (f^{\mu})$, has been analyzed in the mathematical 
literature \cite{LeeVazquez}; it has also been proposed as a phenomenological
''nonextensive'' generalization of classical diffusion \cite%
{PlastinoDiffusion, TsallisDiffusion}. In one dimension and in the absence
of drift, $\overrightarrow{u}\left( \overrightarrow{r},t\right) =0$, this
equation admits of power-law solutions of the form%
\begin{equation}
P\left( r,t;0,0\right) =A\,t^{-\frac{1}{\mu +1}}\left( 1+A^{1-\mu }\frac{%
1-\mu }{\mu \left( \mu +1\right) }\frac{\,r^{2}t^{-\frac{2}{\mu +1}}}{2D}%
\right) ^{\frac{1}{\mu -1}}\,,  
\label{1sol}
\end{equation}%
where the constant $A$ is fixed by normalization. This function can be
written more suggestively as%
\begin{equation}
P\left( r,t;0,0\right) =A\,t^{-\frac{1}{\mu +1}}\,e_{2-\mu }\left( -A^{1-\mu
}\frac{1}{\mu \left( \mu +1\right) }\,\frac{r^{2}t^{-\frac{2}{\mu +1}}}{2D}%
\right) \,,
\label{mu-sol}
\end{equation}%
where the so-called $q$-exponential is defined by%
\begin{equation}
e_{q}\left( x\right) =\left( 1+\left( 1-q\right) x\right) ^{\frac{1}{1-q}}\,,
\end{equation}%
and has the property that $\lim_{q\rightarrow 1}e_{q}\left( x\right) =e^{x}$.

Notice that Eq.(\ref{1}) can be viewed as a classical diffusion equation
with the effective diffusion coefficient $\mu \overleftrightarrow{D}P^{\mu
-1}\left( \overrightarrow{r},t;\overrightarrow{r}_{0},0\right) $. Since
concentration-dependent and density-dependent diffusion coefficients are
commonly used in the scientific literature, this does not differ radically
from classical diffusion: what is unusual is that the effective diffusion
coefficient vanishes when the probability (or concentration) vanishes.

We consider here a collection of atoms moving under Newtonian dynamics and
concentrate on describing the probability of finding a tagged particle, say
the 0-th particle, at position $\overrightarrow{r}$ at time $t$ given it
begins at position $\overrightarrow{r}_{0}$ at time $t=0$,%
\begin{equation}
P\left( \overrightarrow{r},t;\overrightarrow{r}_{0},0\right) =\left\langle
\delta \left( \overrightarrow{q}\left( t\right) -\overrightarrow{r}\right)
\delta \left( \overrightarrow{q}\left( 0\right) -\overrightarrow{r}%
_{0}\right) \right\rangle _{0} \,,  \label{2}
\end{equation}%
where the position of the tagged particle is $\overrightarrow{q}(t)$ and the
notation $\left\langle ...\right\rangle _{0}$ indicates an ensemble average
over some distribution of initial conditions for the other, non-tagged,
particles. Differentiation of Eq.(\ref{2}) with respect to time leads to the
exact equation%
\begin{equation}
\frac{\partial }{\partial t}P\left( \overrightarrow{r},t;\overrightarrow{r}%
_{0},0\right) =\frac{\partial }{\partial \overrightarrow{r}}\cdot 
\overrightarrow{u}\left( \overrightarrow{r},t\right) P\left( \overrightarrow{%
r},t;\overrightarrow{r}_{0},0\right) +\frac{\partial }{\partial 
\overrightarrow{r}}\cdot \overrightarrow{J} \,,  \label{3}
\end{equation}%
with the current%
\begin{equation}
\overrightarrow{J}\left( \overrightarrow{r},t;\overrightarrow{r}%
_{0},0\right) =\left\langle \overrightarrow{v}^{\prime }\left( t\right)
\delta \left( \overrightarrow{q}\left( t\right) -\overrightarrow{r}\right)
\delta \left( \overrightarrow{q}\left( 0\right) -\overrightarrow{r}%
_{0}\right) \right\rangle _{0} \,,  \label{4}
\end{equation}%
where $\overrightarrow{u}\left( \overrightarrow{r},t\right) $ is the average
macroscopic velocity, $\overrightarrow{v}\left( t\right) $ is the
instantaneous velocity of the tagged particle and its peculiar velocity is $%
\overrightarrow{v}^{\prime }\left( t\right) =\overrightarrow{v}\left(
t\right) -\overrightarrow{u}\left( \overrightarrow{r},t\right) $. The idea
of linear response theory is that, aside from convection and in the absence
of driving forces or boundary conditions, the distribution should relax
toward a uniform distribution. Hence, one expects the current to be
proportional, in some sense, to gradients of the distribution and indeed the
simplest assumption that $\overrightarrow{J}\propto \overrightarrow{\nabla }%
P $ gives the usual phenomenological Fick's law. Higher order terms are
expected to involve higher order gradients. Such an expansion is most easily
developed in Fourier space by noting that the Fourier transform of the
current and the distribution are%
\begin{eqnarray}
\widetilde{\overrightarrow{J}}\left( \overrightarrow{k},t;\overrightarrow{r}%
_{0},0\right) &=&\left\langle \overrightarrow{v}^{\prime }\left( t\right)
\exp \left( i\overrightarrow{k}\cdot \overrightarrow{q}\left( t\right)
\right) \delta \left( \overrightarrow{q}\left( 0\right) -\overrightarrow{r}%
_{0}\right) \right\rangle _{0} \,, \\
\widetilde{P}\left( \overrightarrow{k},t;\overrightarrow{r}_{0},0\right)
&=&\left\langle \exp \left( i\overrightarrow{k}\cdot \overrightarrow{q}%
\left( t\right) \right) \delta \left( \overrightarrow{q}\left( 0\right) -%
\overrightarrow{r}_{0}\right) \right\rangle _{0} \,,
\end{eqnarray}%
and expanding their ratio to get a series of the form 
\begin{equation}
\widetilde{\overrightarrow{J}}/\widetilde{P}=i\overrightarrow{k}\cdot 
\overleftrightarrow{D}(t)+... \;,  \label{simple}
\end{equation}%
thus giving the desired gradient expansion \cite{DuftyDiffusion}. In
general, the diffusion coefficient, which is given by an Einstein relation,
is time dependent on microscopic time scales characterized by the time
between atomic collisions, but in the long-time limit appropriate to
hydrodynamics it tends toward a constant value.

Comparison of Eqs.(\ref{1}) and (\ref{3}) suggests that the key assumption
in the use of the generalized diffusion equation is that the current is no
longer most simply described by gradients in the distribution function, but
rather by gradients in the distribution raised to some power. In order to
generalize the linear-response argument, we proceed by defining the Fourier
transform of the $\mu $-th power of the distribution to be 
\begin{equation}
\widetilde{P}_{\mu }\left( \overrightarrow{k},t;\overrightarrow{r}%
_{0},0\right) =\int d\overrightarrow{r}\;\exp \left( i\overrightarrow{k}%
\cdot \overrightarrow{r}\right) P^{\mu }\left( \overrightarrow{r},t;%
\overrightarrow{r}_{0},0\right) \,,  \label{power}
\end{equation}%
and expand in the wavevector to obtain 
\begin{equation}
\widetilde{\overrightarrow{J}}/\widetilde{P}_{\mu }=i\overrightarrow{k}\cdot 
\overleftrightarrow{D}_{\mu }+...\;,  \label{expansion}
\end{equation}%
which gives 
\begin{eqnarray}
\overleftrightarrow{D}_{\mu }\left( t\right) &=&\left\langle \overrightarrow{%
q}\left( t\right) \overrightarrow{v}^{\prime }\left( t\right) \delta \left( 
\overrightarrow{q}\left( 0\right) -\overrightarrow{r}_{0}\right)
\right\rangle _{0}
\left[ \int d\overrightarrow{r}\;P^{\mu }\left( \overrightarrow{r%
},t;\overrightarrow{r}_{0},0\right) \right]^{-1}  \notag  \\
&=&\overleftrightarrow{D}_{1}\left( t\right) 
\left[ \int d\overrightarrow{r}%
\;P^{\mu }\left( \overrightarrow{r},t;\overrightarrow{r}_{0},0\right) \right]^{-1} \,,
\label{einstein}
\end{eqnarray}%
where $\overleftrightarrow{D}_{1}\left( t\right) $ is the classical
diffusion coefficient which is the same as that occurring in Eq.(\ref{simple}).
By expressing $\overrightarrow{q}\left( t\right) $ in terms of
the time-integral of the velocity, the diffusion coefficient becomes a
function of the velocity auto-correlation function as expected%
\begin{equation}
\overleftrightarrow{D}_{\mu }\left( t\right) =\int_{0}^{t}\left\langle 
\overrightarrow{v}\left( t^{\prime }\right) \overrightarrow{v}^{\prime
}\left( t\right) \delta \left( \overrightarrow{q}\left( 0\right) -%
\overrightarrow{r}_{0}\right) \right\rangle _{0}dt^{\prime }
\left[ \int d \overrightarrow{r}\;P^{\mu }\left( \overrightarrow{r},t;\overrightarrow{r}%
_{0},0\right) \right]^{-1} \,.  
\label{gen_einstein}
\end{equation}%
Then, keeping only the lowest order in (\ref{expansion}), we obtain the
generalized advection-diffusion equation 
\begin{equation}
\frac{\partial }{\partial t}P\left( \overrightarrow{r},t;\overrightarrow{r}%
_{0},0\right) =\frac{\partial }{\partial \overrightarrow{r}}\cdot 
\overrightarrow{u}\left( \overrightarrow{r},t\right) P\left( \overrightarrow{%
r},t;\overrightarrow{r}_{0},0\right) \,+\,\frac{\partial }{\partial 
\overrightarrow{r}}\cdot \overleftrightarrow{D}_{\mu }\left( t\right) \cdot 
\frac{\partial }{\partial \overrightarrow{r}}P^{\mu }\left( \overrightarrow{r%
},t;\overrightarrow{r}_{0},0\right) \,,  \label{q-diffusion}
\end{equation}%
with a generalized Einstein relation, Eq.(\ref{gen_einstein}), relating the
diffusion coefficient to the underlying microscopic dynamics and
corresponding distribution function. Note that even on hydrodynamic time
scales, for which $\overleftrightarrow{D}_{1}\left( t\right) $ can be
replaced by $\overleftrightarrow{D}_{1}=\lim_{t\rightarrow \infty }%
\overleftrightarrow{D}_{1}\left( t\right) $, $\overleftrightarrow{D}_{\mu
}\left( t\right) $ remains time-dependent due to its dependence on $P^{\mu
}\left( \overrightarrow{r},t;\overrightarrow{r}_{0},0\right) $.

Scaling solutions of Eq.(\ref{q-diffusion}) in the absence of drift ($%
\overrightarrow{u}=0$) are easily examined. For simplicity, consider the
approximation, expected to be valid on hydrodynamic time scales, $%
\overleftrightarrow{D}_{1}\left( t\right) \sim $ $\lim_{t\rightarrow \infty }%
\overleftrightarrow{D}_{1}\left( t\right) \equiv \overleftrightarrow{D}_{1}$%
, and suppose that in $d$-dimensions the distribution assumes the form $%
P\left( \overrightarrow{r},t;\overrightarrow{r}_{0},0\right)
=t^{-da}F(r/t^{a})$ for some exponent $a$ (note that the prefactor is
dictated by normalization). Then, one has 
\begin{eqnarray}
\int d\overrightarrow{r}\;P^{\mu }\left( \overrightarrow{r},t;%
\overrightarrow{r}_{0},0\right)  &=&\int d\overrightarrow{r}\;t^{-d\mu
a}F^{\mu }(r/t^{a})  \notag \\
&=&S_{d}t^{-d\left( \mu -1\right) a}\int_{0}^{\infty }F^{\mu }(x)x^{d-1}dx\,,
\end{eqnarray}%
with ${x}\equiv t^{-a}|\overrightarrow{r}|$ and 
where $S_{d}$ is the area of the $d-$dimensional hypersphere. The
generalized diffusion coefficient is then%
\begin{equation}
\overleftrightarrow{D}_{\mu }\left( t\right) =t^{d\left( \mu -1\right) a}%
\overleftrightarrow{D}_{\mu }^{\prime }\,,
\end{equation}%
where%
\begin{equation}
\overleftrightarrow{D}_{\mu }^{\prime } \equiv \overleftrightarrow{D}_{1}\,
\left[ S_{d}\int_{0}^{\infty} F^{\mu }(x)\,x^{d-1}dx\right] ^{-1}
\end{equation}%
is a time-independent coefficient. Substitution into Eq.(\ref{q-diffusion})
gives, after simplification, 
\begin{equation}
-a \left( d\,F(x)+ \overrightarrow{x} \cdot
 \frac {\partial} {\partial \overrightarrow{x}} F(x)\right) =t^{1-2a}\,%
\frac{\partial }{\partial \overrightarrow{x}}\cdot \overleftrightarrow{D}%
_{\mu }^{\prime }\cdot \frac{\partial }{\partial \overrightarrow{x}}F^{\mu
}\left( x\right) ,
\end{equation}%
which, for scaling consistency,  is only valid for $a=\frac{1}{2}$ .  
The generalized diffusion equation given here therefore describes normal
diffusion when there is no drift. However, the distributions are not Gaussian. 
For example, in one dimension the scaling equation becomes%
\begin{equation}
-\frac{1}{2}F(x)-\frac{1}{2}x\frac{d}{dx}F(x)\,=\,D_{\mu }^{\,\prime }\frac{%
d^{2}}{dx^{2}}F^{\mu }\left( x\right) \,,
\end{equation}%
which is again solved by the canonical $q$-exponential form as 
\begin{equation}
F\left( x\right) =A\, t^{-1/2}\, e_{2-\mu }\left( -\frac{A^{1-\mu }}{\mu \left( \mu
+1\right) }\,\frac {x^{2}}{2 D\,_{\mu }^{\prime}} \right) \,,  
\label{sol_mu}
\end{equation}%
or, with $\mu =2-q$, 
\begin{equation}
P\left( r,t;0,0\right) =A\,t^{-1/2}\left( 1-\left( 1-q\right) \frac{A^{q-1}}{%
\left( 2-q\right) \left( 3-q\right) }\,\frac{r^{2}}{2D\,_{\mu }^{\prime }t}%
\right) ^{\frac{1}{1-q}}\,.  \label{sol_q}
\end{equation}%

The mean-squared displacement is then obtained straightforwardly from (\ref%
{sol_q}); with $r^{\ast }=r/\sqrt{D\,_{\mu }^{\prime }t}$, one has 
\begin{eqnarray}
\langle r^{2}(t)\rangle  &=&\int_{0}^{\infty }dr\,r^{2}\,P\left(
r,t;0,0\right)   \notag \\
&=&D\,_{\mu }^{\prime }\,t\,\int_{0}^{\infty }dr^{\ast }\,r^{\ast 2}A\left(
1-\left( 1-q\right) \frac{A^{q-1}}{\left( 2-q\right) \left( 3-q\right) }%
\frac{r^{\ast 2}}{2}\right) ^{\frac{1}{1-q}}\,.  \label{msqrd}
\end{eqnarray}%
Thus, in the absence of drift, Eq.(\ref{q-diffusion}) describes normal
diffusion in the sense that the mean-squared displacement grows linearly
with time, but it generalizes classical diffusion as the distribution (\ref%
{sol_q}) takes on the form typical of nonextensive systems.

Although Eq.(\ref{q-diffusion}) is formally the same as Eq.(\ref{1}), there
is an additional self-consistency implied by the Einstein relation (\ref%
{gen_einstein}) which raises the question as to whether or not these
descriptions can be regarded in any sense as being the same. In some cases
they are, because Eq.(\ref{q-diffusion}) is of first order in the time so
that any time-dependence generated by the self-consistency can be hidden by
a non-linear change in the time variable. For example, in the absence of
drift, let $Q\left( \overrightarrow{r},s;\overrightarrow{r}_{0},0\right) $
be a solution to Eq.(\ref{1}), written with the variable $s$ instead of $t$,
and having initial condition $P\left( \overrightarrow{r},0;\overrightarrow{%
r}_{0},0\right) $. Then it is easy to see that the solution of Eq.(\ref%
{q-diffusion}) is $P\left( \overrightarrow{r},t;\overrightarrow{r}%
_{0},0\right) =Q\left( \overrightarrow{r},s\left( t\right) ;\overrightarrow{r%
}_{0},0\right) $ where the time variables are related by the implicit
equation%
\begin{equation}
t\left( s\right) =\int_{0}^{s}\left( \int d\overrightarrow{r}\;Q^{\mu
}\left( \overrightarrow{r},\sigma ;\overrightarrow{r}_{0},0\right) \right)
d\sigma \,.  \label{time}
\end{equation}%
To illustrate, consider again the case of one dimensional diffusion with no
drift term. The solution of the porous media equation (\ref{1}) is given by
(\ref{mu-sol}) which, with the replacement $t\rightarrow s$, is what is called
here $Q$. Substitution into Eq.(\ref{time}) gives%
\begin{equation}
t\left( s\right) =\int_{0}^{s}\left( \int d\overrightarrow{r}\;Q^{\mu
}\left( \overrightarrow{r},\sigma ;\overrightarrow{r}_{0},0\right) \right)
d\sigma =Cs^{\frac{2}{\mu +1}}\,,
\end{equation}%
for some constant $C$ , thus recovering Eq.(\ref{sol_mu}).

It is interesting that Eq.(\ref{q-diffusion}) is very similar to the
equation recently derived by Abe and Thurner \cite{AbeThurner} where the
derivation begins with the classical picture based on a random walker which
jumps from position $r+\Delta $ at time $t$ to position $r$ at time $t+\tau $
with a jump probability $\phi \left( \Delta \right) = \phi \left( -\Delta
\right)$, so that the probability that the walker be at $r$ at time $t+\tau$
is given by%
\begin{equation}
{\cal P} \left( r,t+\tau \right) =\int {\cal P} \left( r+\Delta ,t\right) \phi \left( \Delta
\right) \;d\Delta \,.
\end{equation}%
They consider the more general case in which the probability ${\cal P} \left(
r,t\right) $ occuring in the integral is replaced by the so-called escort
probability 
\begin{equation}
{\cal F}\left( r,t\right) =\frac{{\cal P}^{\nu }\left( r,t\right) } 
{\int {\cal P}^{\nu} \left( r,t \right) dr} \;,
\end{equation}%
with the result that the diffusion equation becomes%
\begin{equation}
\frac{\partial }{\partial t}{\cal P} \left( r,t\right) =D_{\nu }\left( t\right) 
\frac{\partial ^{2}}{\partial r^{2}}{\cal P}^{\nu }\left( \overrightarrow{r}%
,t\right) +\frac{1}{\tau }\left( {\cal F} \left( r,t\right) -{\cal P}\left( \overrightarrow{%
r},t;\right) \right)\,,  \label{nu-diff-eq}
\end{equation}%
with $D_{\nu }\left( t\right) $ defined as in Eq.(\ref{gen_einstein}). 
Equation (\ref{nu-diff-eq}) is somewhat difficult to interpret as its
derivation involves taking the limit $\tau \rightarrow 0$ in which case the
second term on the right is problematic. Apart from this problem, the result is
strikingly similar to our nonlinear-response result including the definition
of the time-dependent diffusion coefficient (see Eq.(15) in \cite{AbeThurner}).
Comparison of the results also indicates that ``nonextensive'' expressions 
can arise without the explicit introduction of escort probabilities.

In this paper, we have developed a method whereby a statistical mechanical 
derivation yields typically nonextensive expressions by generalizing a key ansatz,
i.e. a generalization of standard linear response arguments which gives rise
to a $q$-diffusion equation including a generalization of the Einstein
relation for the diffusion coefficient. The equation is similar in structure
to the porous media equation, but with the important difference that the
diffusion coefficient depends on the solution of the equation which leads to
the fact that, in the absence of drift, the diffusion process is classical
with mean-squared displacement increasing linearly with time.
The new $q$-diffusion equation admits of the well known $q$%
-exponential distribution which is often postulated as the signature of
nonequilibrium systems.

\end{document}